\documentclass[traditabstract, onecolumn]{aa_mod}
\usepackage{graphicx, natbib, txfonts, hyperref}
\bibpunct{(}{)}{;}{a}{}{,} 

\begin{document}

\title{\textbf{\emph{Scanamorphos}} for the APEX-ArT\'eMiS 350-450\,$\mu$m camera : description and user guide}

\author{H. Roussel
}

\institute{Institut d'Astrophysique de Paris, Universit\'e Pierre et Marie Curie (UPMC), Sorbonne Universit\'e,
CNRS (UMR 7095), \\
75014 Paris, France\label{inst1} \\
\email{roussel@iap.fr} \\
~ \\
software webpage:~ \url{http://www2.iap.fr/users/roussel/artemis/} \\
~ \\
ESO-APEX portal:~ \url{http://www.apex-telescope.org/instruments/pi/artemis/} \\
~ \\
version: v3.1 (first public delivery: v3 on March 8, 2018) \\
~ \\
members of the ArT\'eMiS team who have helped interfacing with the pipeline and doing tests: \\
Guillaume Willmann, Fr\'ed\'eric Schuller, Philippe Andr\'e, Pascal Gallais (CEA Saclay, AIM) \\
~ \\
{\bf If you are using this software, please cite both this document (as well as the pipeline documentation,
accessible through the ESO-APEX portal, in the Data Reduction section) and the original refereed paper describing
\emph{Scanamorphos} for the PACS and SPIRE \emph{Herschel} photometers:}~~ Roussel, H. 2013, PASP, 125, 1126\,.
}

\abstract{
{\it Scanamorphos} is public software initially developed to post-process scan observations
performed with the {\it Herschel} photometer arrays. This post-processing mainly consists in subtracting
the total low-frequency noise (both its thermal and non-thermal components), masking cosmic ray hits,
and projecting the data onto a map. Building upon the results obtained for P-ArT\'eMiS (the prototype
of ArT\'eMiS), {\it Herschel} and then NIKA2 (a resident camera of the IRAM 30-m telescope operating
at 1.25 and 2\,mm), it has now been tailored to the ArT\'eMiS camera, an ESO and OSO P.I. instrument
installed at the APEX 12-m telescope, demonstrating our initial claim that the software principles were
directly transposable to scan observations made with other instruments, including from the ground,
provided they entail sufficient redundancy. This document explains how the algorithm was modified
to cope with the specificities of ArT\'eMiS observations and with the atmospheric emission at 350
and 450\,$\mu$m, far dominating the instrumental drifts that were the only low-frequency noise
component in {\it Herschel} data. Like in the original software, this was accomplished without assuming
any noise model and without applying any Fourier-space filtering, by exploiting the redundancy built
in the observations -- taking advantage of the fact that each portion of the sky is sampled at multiple
times by multiple bolometers. It remains an interactive software in the sense that the user is
allowed to optionally visualize and control results at each intermediate step, but the processing
is fully automated. It has been grafted onto the ArT\'eMiS pipeline, in charge of the formatting,
calibration and projection of the data, that is described elsewhere.
}

\titlerunning{{\it Scanamorphos for ArT\'eMiS}}

\maketitle

This document serves as an introduction and user guide to {\it Scanamorphos} tailored to ArT\'eMiS.
For a thorough presentation of the ArT\'eMiS camera, see for exemple \cite{Reveret14}. 
It employs the same technology as the PACS photometers \citep{Billot06}, namely filled bolometer arrays,
and the same multiplexing architecture, so the change of instrument is almost seamless. The array
geometry has changed with time (the number of subarrays went from 4 to 8), but this is easily
accounted for, and observations combining different instrument configurations can be simultaneously
processed. The main changes brought along by the ArT\'eMiS data are first the presence of the atmospheric
emission, behaving as an intense and fast-varying correlated low-frequency noise, and second
the observation strategy and scan characteristics. Since the software principles and initial algorithm
for {\it Herschel} were described at length in the companion paper \citep{Roussel13}, we here
focus on the main modifications made necessary in the context of observations on APEX. It is
assumed that the definitions and algorithm steps given in that paper do not need to be recalled.
The first array to have been commissioned being that operating at 350\,$\mu$m, illustrations
are based on this sole array. We have used the mosaic of the NGC\,6334 star formation region.
The ArT\'eMiS data processed with an earlier version of {\it Scanamorphos} have been analyzed and
published by \cite{Andre16}. The {\it Herschel} photometric data from the HOBYS program that were used
to characterize the large spatial scales in that paper have been presented by \cite{Russeil13}
and \cite{Tige17}.

\section{Specificities of ArT\'eMiS observations on APEX}

\begin{figure}[ht!]
\centering
\begin{minipage}[!ht]{7.cm}
\includegraphics[width=1.\textwidth,clip]{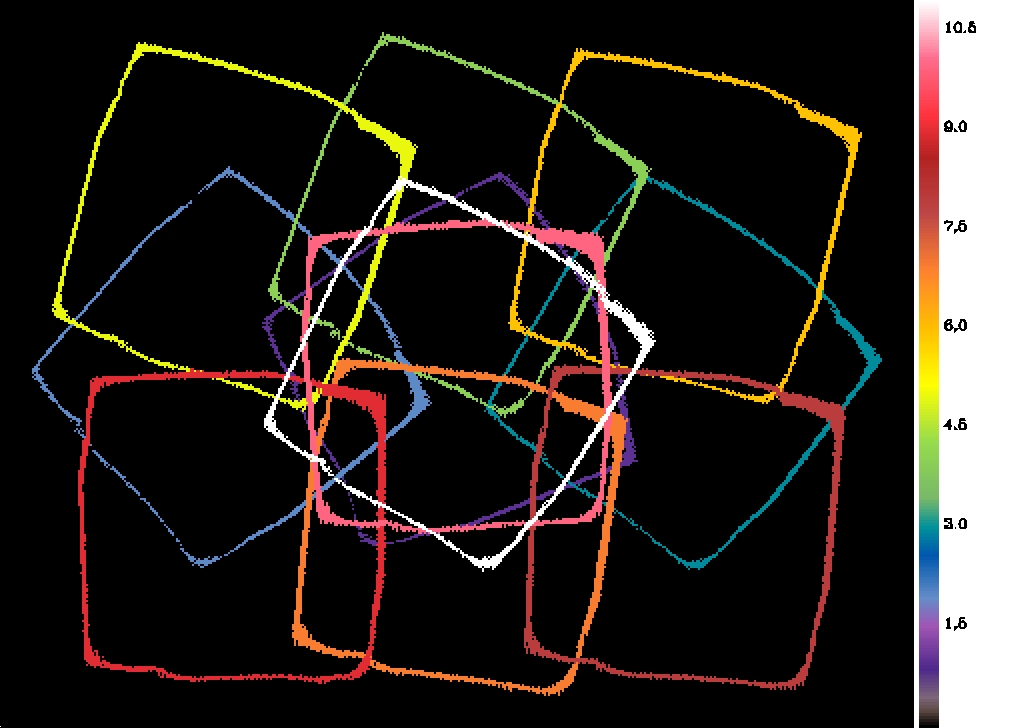}
\end{minipage}
\begin{minipage}[!ht]{7.cm}
\includegraphics[width=1.\textwidth,clip]{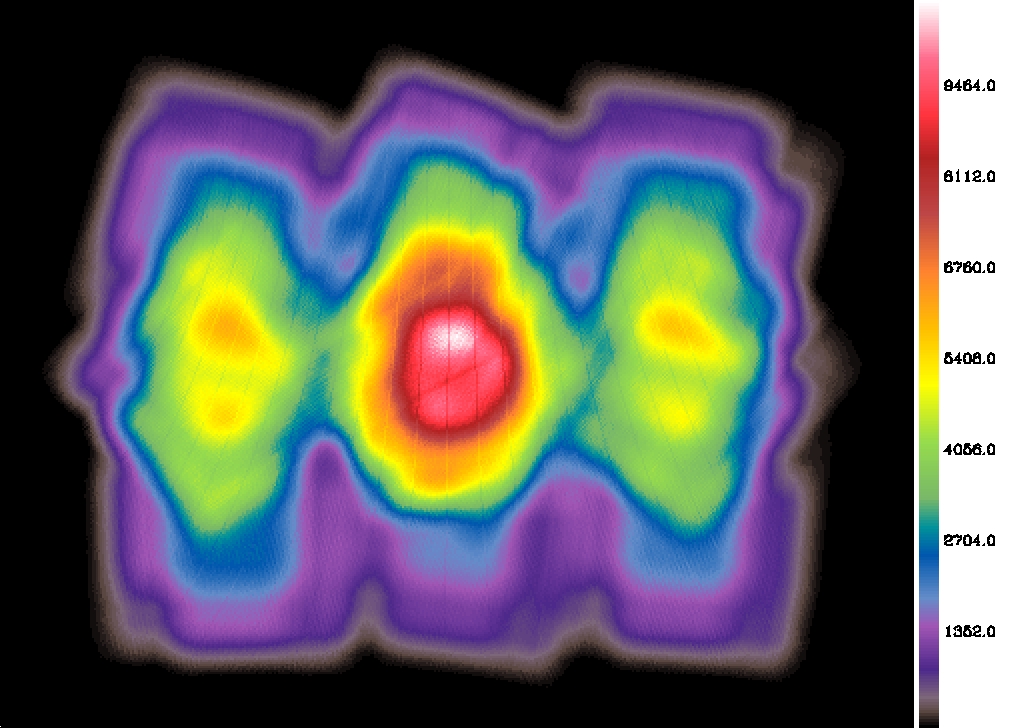}
\end{minipage}
\vspace*{0.1cm} \\
\begin{minipage}[!ht]{7.cm}
\includegraphics[width=1.\textwidth,clip]{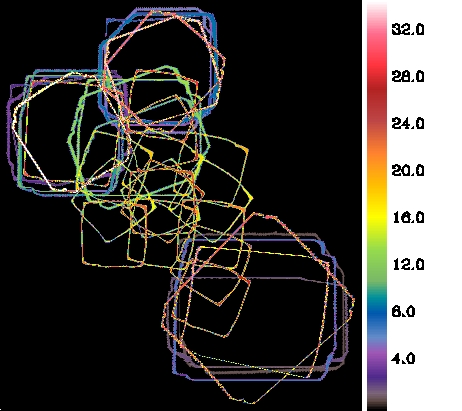}
\end{minipage}
\begin{minipage}[!ht]{7.cm}
\includegraphics[width=1.\textwidth,clip]{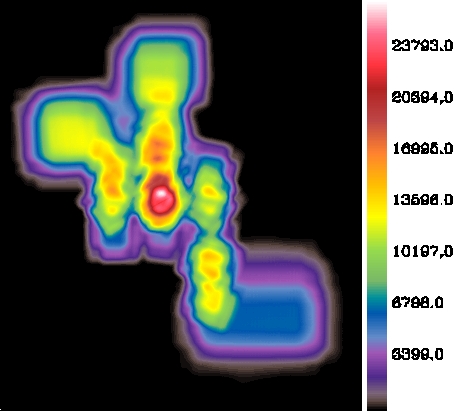}
\end{minipage} \\
\caption{{\bf Top left:} Outline of the 11 scans composing the central part of the NGC\,6334 mosaic.
{\bf Top right:} Corresponding weight map. {\bf Bottom:} Same as above for the 35 scans of the
whole mosaic, extending the field of view to the north, north-east and south-west.}
\label{fig:outline}
\end{figure}

ArT\'eMiS observations of star formation regions consist in extended mosaics of small maps
(Fig.\,\ref{fig:outline}), in which the coverage is very inhomogeneous and the noise even more (due to the
fact that the atmospheric conditions can vary a lot between different scans taken hours or days
apart). This is unavoidable, the maximum duration of a scan and the maximum scan velocity being
both constrained. It is thus difficult to plan mosaics such that each region is covered by at least
two scans with two distinct position angles, as close to being orthogonal as practical.
In practice, regions obeying this requirement constitute only a fraction of the mosaic.
Another important change is that the separation between scan legs is very small (of the order
of the size of a detector). This implies that the data volume per solid angle is increased
by a large factor with respect to {\it Herschel} observations.

Because the individual maps are small, and because the redundancy at distinct
scan position angles can be very limited, it is not possible to recover very extended emission
(i.e. more extended than individual maps). But we insist that if these maps got larger than the
detector array, and provided each location were observed with at least two distinct scan orientations,
then nothing should prevent scales larger than the array to be recovered.

We also have to account for the fact that the scan velocity is less stable at a ground-based telescope
than in space observations. The relative dispersion is typically of the order of 15\% (rms).
This matters, because the median scan velocity and the sampling rate (that is fixed) control the
stability length, i.e. the length over which the drifts will be considered stable
\citep[see Section 3.2 in][]{Roussel13}. This parameter naturally has to be fixed for the processing
of the whole observations. A certain variability thus has to allowed in the fiducial number of samples
per stability length.

A so-called ``spiral mode'' is enabled at the APEX telescope, in which the scans cannot be sliced into
distinct scan legs because the scan orientation is continuously and smoothly changing. Another
consequence is that the running average of the scan velocity can no longer be considered constant
within a single scan. For the algorithm, we expect that the only step to be modified accordingly
is the baseline subtraction, since it currently relies on the slicing into scan legs. The rest
of the algorithm should be unaffected, except for a few parameters to be adjusted. The option
to process ``spiral-mode'' observations is not enabled yet, but may be in the future (on a
best-effort basis, provided this mode is regularly used for science observations).

\section{Atmospheric emission}

\begin{figure}[ht!]
\vspace*{0.3cm}
\centering
\begin{minipage}[!ht]{7.cm}
\includegraphics[width=1.\textwidth,clip]{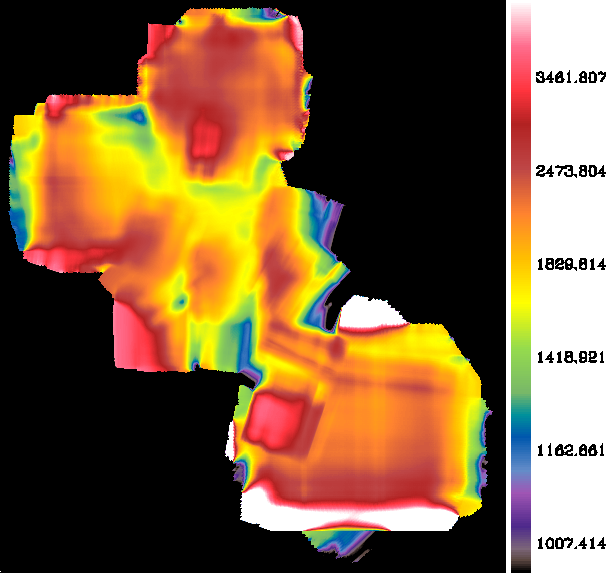}
\end{minipage}
\begin{minipage}[!ht]{7.cm}
\includegraphics[width=1.\textwidth,clip]{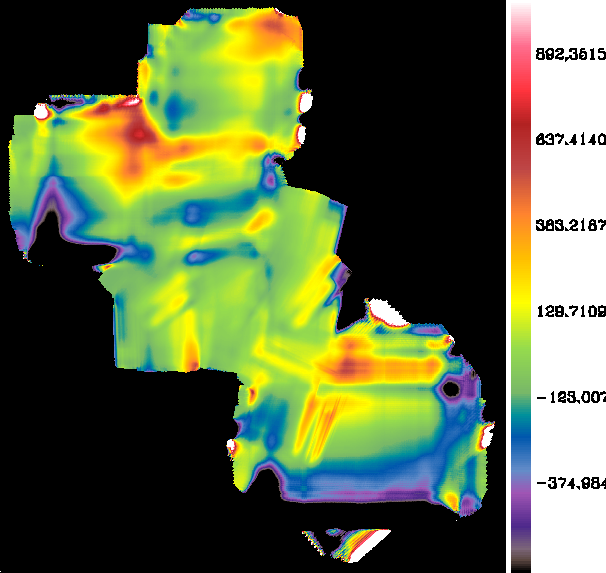}
\end{minipage}
\vspace*{0.1cm} \\
\begin{minipage}[!ht]{7.cm}
\includegraphics[width=1.\textwidth,clip]{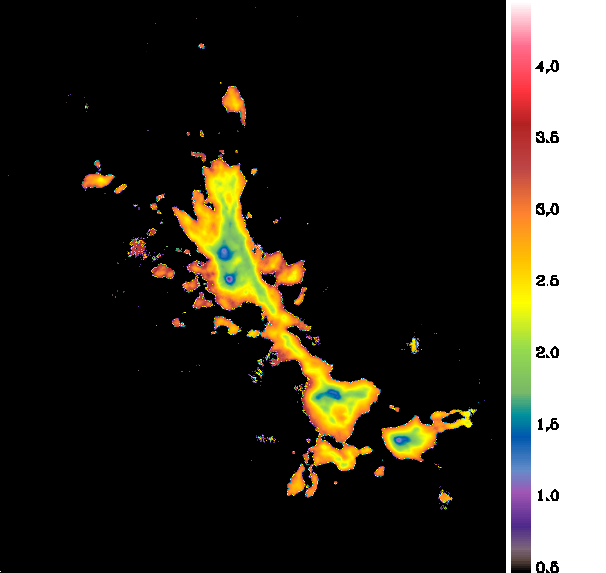}
\end{minipage}
\begin{minipage}[!ht]{7.cm}
\includegraphics[width=1.\textwidth,clip]{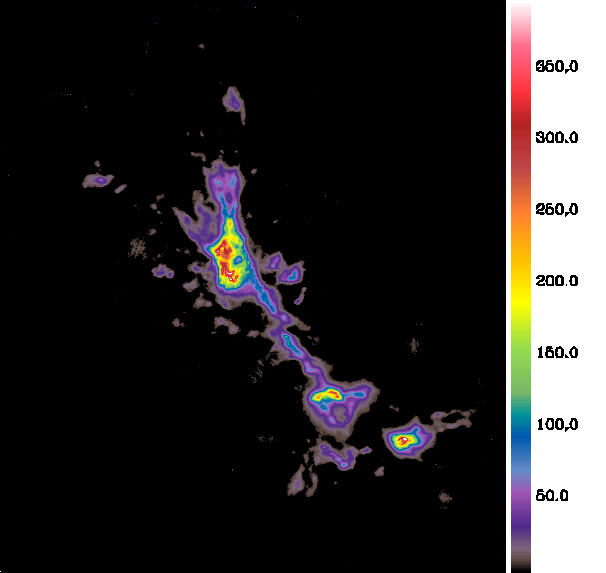}
\end{minipage} \\
\caption{Background and noise levels. {\bf (1)} Average drift map (with global offsets in each scan to ensure
positivity). {\bf (2)} Individual drifts map. {\bf (3)}~Logarithm of the ratio of the average drift to the signal
above $6\, \sigma$. {\bf (4)} Final signal to noise ratio map (above $6\, \sigma$).}
\label{fig:atmo}
\end{figure}

The ArT\'eMiS camera does not allow the measurement of the absolute brightness (just like PACS and SPIRE),
implying that only differential measurements can be discussed. In addition, large-scale gradients are
also erased by the observation strategy and the processing. Even with these restrictions,
the atmospheric emission measured in the ArT\'eMiS observations at 350 to 450\,$\mu$m is brighter than
any astronomical signal in the Galactic disk by several orders of magnitude. This is illustrated in
Figure\,\ref{fig:atmo}, showing both the average component (overwhelmingly dominated by the atmosphere)
and the complementary part of the subtracted drifts, compared with the signal of NGC\,6334 at 350\,$\mu$m.

\section{Main modifications to the algorithm}

The main algorithmic changes stem both from the atmosphere brightness and variability
on relatively short timescales, and from the observation strategy causing the intra-scan
redundancy to be high but the inter-scan redundancy to be very low. \\
1) High-frequency noise: \\
The high-frequency noise (used to compute weights) can vary significantly within a single scan
if the atmosphere is not perfectly stable. It is thus computed for each scan leg, except for
short scans where sources are dominant, in which case it is estimated for the whole scan.
The same applies for the so-called threshold noise. \\
2) Unstable detectors: \\
PACS and ArT\'eMiS detectors can oscillate between high and low states. Instead of detecting
and masking time intervals affected by brightness jumps as for PACS, which would not be practical
for ArT\'eMiS data, the few most unstable detectors are identified through the bimodality of their
brightness distribution in most legs of a given scan, and entirely discarded. \\
3) Baseline subtraction: \\
The construction of the source mask (see next section) is complicated by the fact that different
scans have largely disjoint fields of view and that bright sources occupy a vastly varying fraction
of the area of each individual map. For small scans in particular, special care is needed to determine
the background level and average noise, setting the threshold level for the mask. The mask is
thus built scan per scan before being merged. \\
4) Destriping: \\
The destriping module has to be deactivated, mainly because the large fluctuations of the
atmosphere preclude linear fits of the differential signal (i.e. the difference between the signal
of a given detector in a given scan leg and the signal simulated for the same trajectory from all
the data). This is especially true when scan legs are short and cover a minute fraction of
the field extent. \\
5) Average drift subtraction: \\
The computation of the average drift produces a spurious component that we called the excess
average drift \citep[see Section 3.5.2 in][]{Roussel13}. We found that it is far better removed
by another iteration of the baseline subtraction (Section 3.4) than by the method described
in the paper, that was adapted to observations with suitable inter-scan redundancy throughout
the field of view. Furthermore, the atmosphere is much more prone to have a high degree of repeatability
in successive scans than the instrumental thermal drift, which is an additional reason to drop
the original method. \\
6) Additional components of correlated noise: \\
We initially assumed that, beyond the correlated noise of the full array, there were
additional components of correlated noise, specific to each subarray (of $16 \times 18$ detectors),
and subtracted on the same successively shorter timescales as the uncorrelated noise
\citep[see Section 3.5.5 in][]{Roussel13}. For the tested fields, however, this was not found
to bring any improvement, so this correction was suppressed. Other correlation patterns may be
present in the data; this will be explored in the future.
\\

\section{Input data and parameters}

Input data are in the form of structures, one for each scan. The data at 350 and 450\,$\mu$m
are stored in separate structures and have to be processed separately, because of the large
data volume.
A utility is provided to convert the pipeline output files, containing data that are
corrected for the opacity and calibrated (in the format produced by the IDL ``save'' routine,
with the ``.xdr'' extension) to {\it Scanamorphos} input data. The pipeline outputs one
file per scan leg, instead of one file per scan. This is easily accounted for, since all
the files corresponding to a single scan are given names sharing a unique root. The same
utility also converts the time series back to the pipeline format at the end of the
processing, for the final projection.
An example script to reduce raw data up to level 1 (by analogy with {\it Herschel}) with the
pipeline and then up to level 2 with {\it Scanamorphos} is provided with the pipeline distribution.
The latter contains both {\it Scanamorphos} routines per se and interfaces from and back to
the pipeline. \\

\begin{figure}[ht!]
\vspace*{-0.2cm}
\centering
\begin{minipage}[!ht]{5.cm}
\includegraphics[width=1.\textwidth,clip]{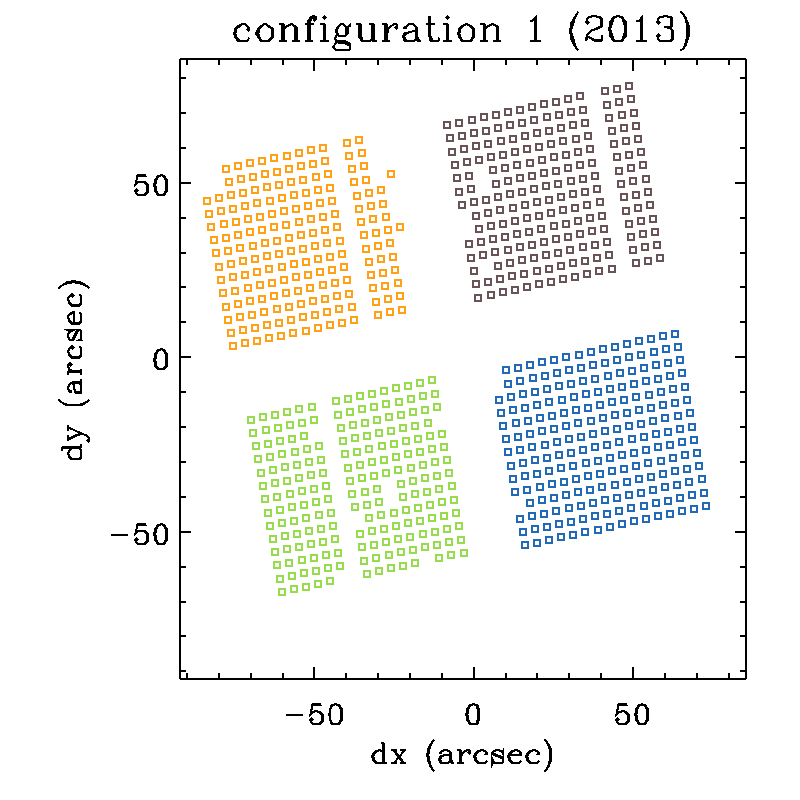}
\end{minipage}
\begin{minipage}[!ht]{5.cm}
\includegraphics[width=1.\textwidth,clip]{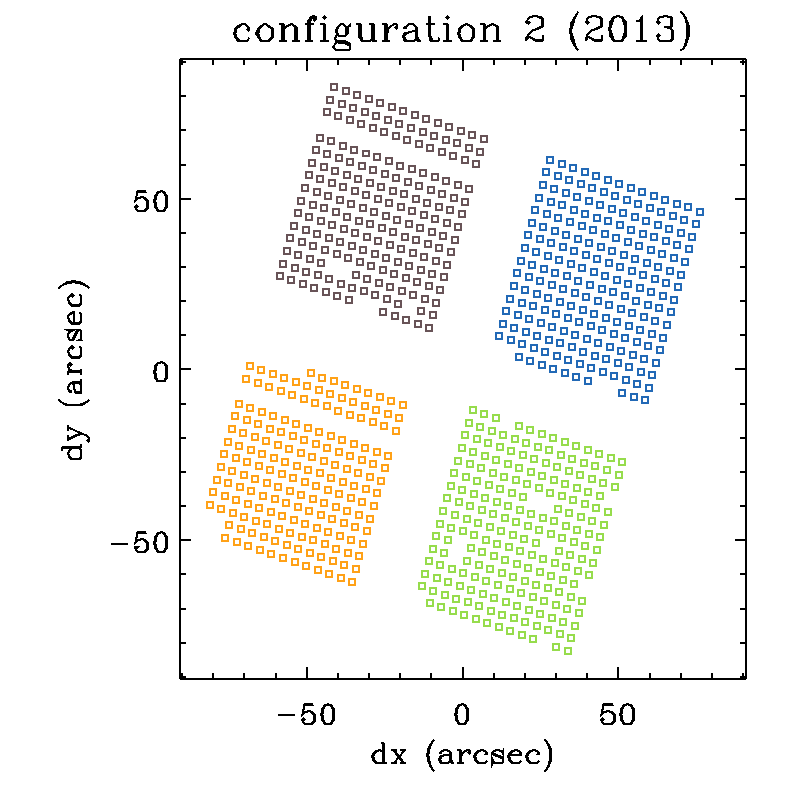}
\end{minipage}
\begin{minipage}[!ht]{6.cm}
\includegraphics[width=1.\textwidth,clip]{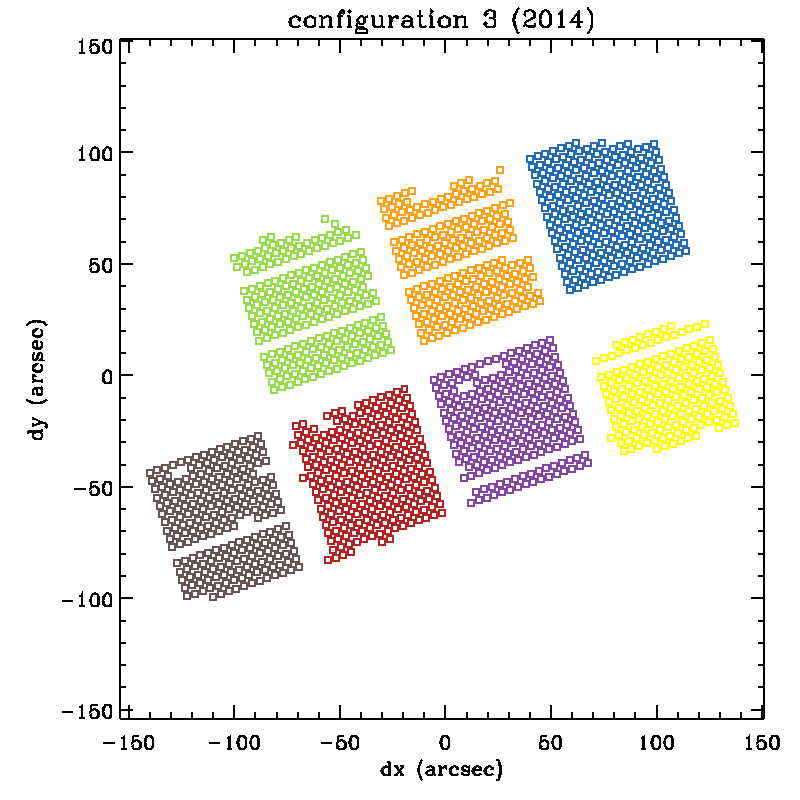}
\end{minipage} \\
\caption{Visualization of the different array geometries of the NGC\,6334 observations.
Each subarray is shown with a distinct color at the start of the first scan in each of the three
configurations (with exact relative map coordinates but detectors not to scale).
\vspace*{-0.5cm}
~
}
\label{fig:geom}
\end{figure}

\begin{figure}[ht!]
\vspace*{0.5cm}
\centering
\begin{minipage}[!ht]{8.cm}
\includegraphics[width=1.\textwidth,clip]{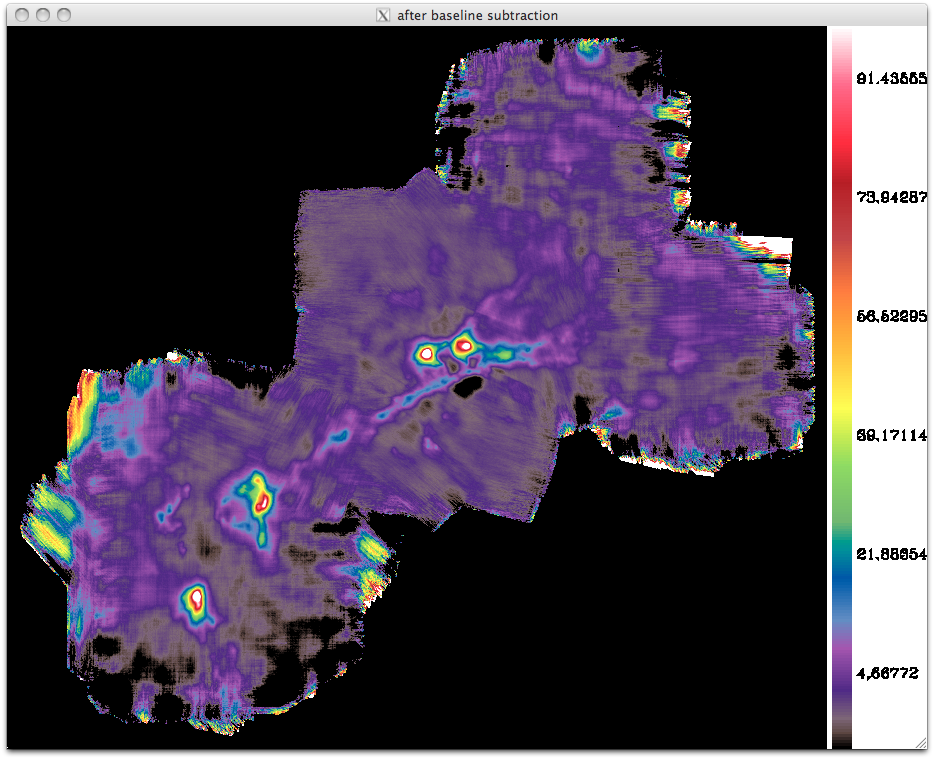}
\end{minipage}
\begin{minipage}[!ht]{8.cm}
\includegraphics[width=1.\textwidth,clip]{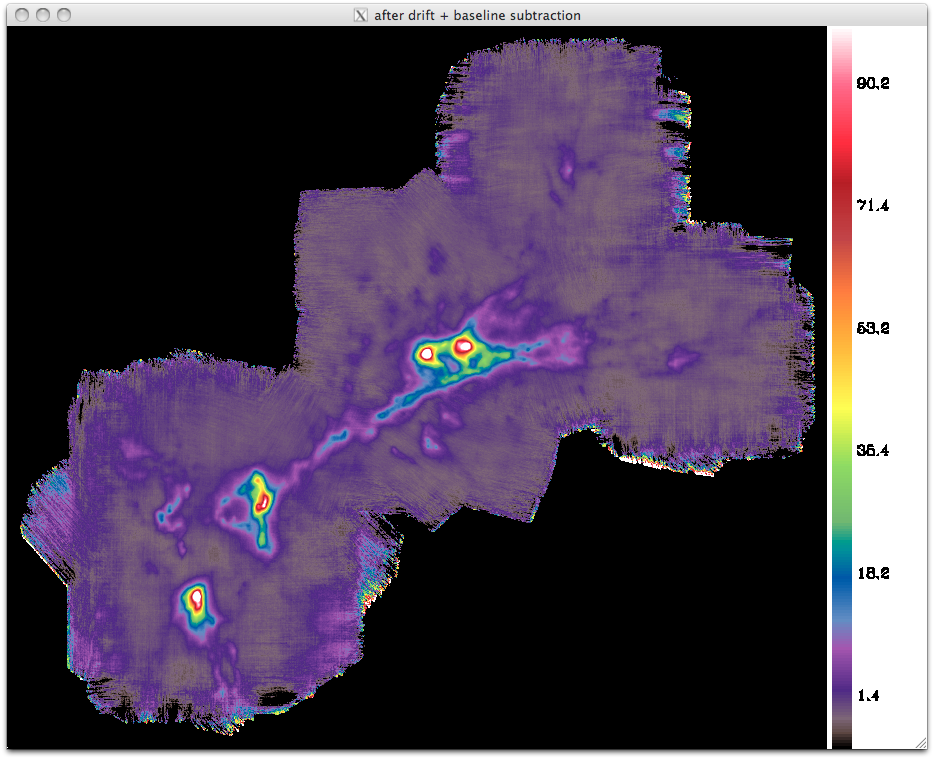}
\end{minipage}
\vspace*{0.1cm} \\
\begin{minipage}[!ht]{8.cm}
\includegraphics[width=1.\textwidth,clip]{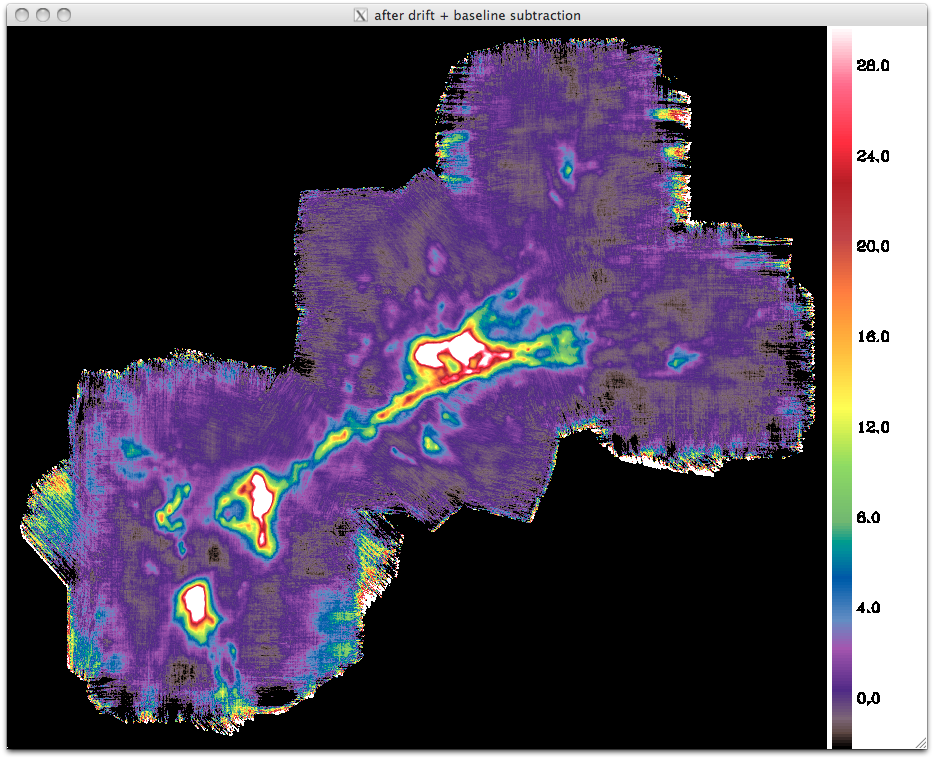}
\end{minipage}
\begin{minipage}[!ht]{8.cm}
\includegraphics[width=1.\textwidth,clip]{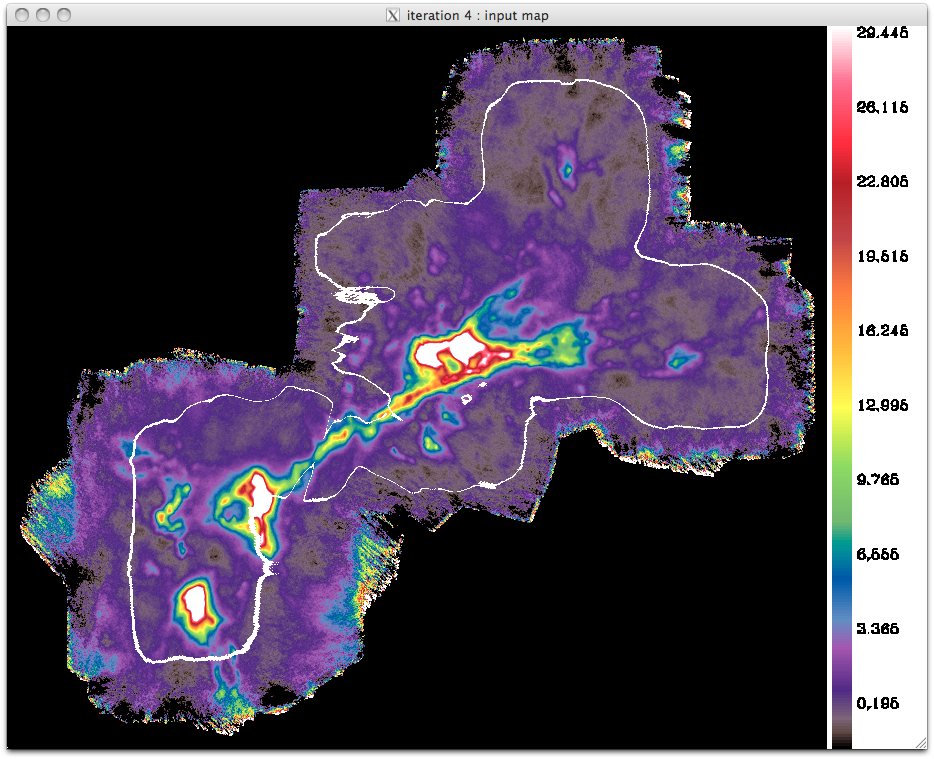}
\end{minipage} \\
\caption{Maps projected after some of the main processing steps (when using\, {\tt /visu}\,).
{\bf (1)} After the initial baseline subtraction, removing noise on the longest timescales. Bright sources
are not surrounded by negative bowls, such as introduced by a highpass filter, but at this stage most scans
still contain a lot of noise on intermediate scales. {\bf (2)} After the average drift subtraction
(on small timescales) and a new iteration of the baseline subtraction, displayed on the same brightness
range as (1). {\bf (3)} Same as (2), but displayed so as to emphasize low-brighness regions.
{\bf (4)} Nearing the end of the processing, displayed on the same brightness range as (3). The added white
contour indicates regions with the best coverage (see Fig.~\ref{fig:outline}). Outside, the residual correlated
noise is significantly higher.
\vspace*{-0.2cm}
~
}
\label{fig:processing}
\end{figure}

The {\tt /visu} option allows to pause after each important processing step and visualize
intermediate results; the processing is resumed by clicking on a dialog box. With respect
to the original code, the visualization of the geometry of the subarrays is added (Fig.~\ref{fig:geom}).
Figure~\ref{fig:processing} shows a few of the intermediate maps displayed in interactive mode.
The imprint and orientation of the scans can be visually checked with the {\tt /vis\_traject}
option. The {\tt /debug} option is in principle useful only for the developer.

The {\tt /galactic} option has a slightly different meaning for ArT\'eMiS and for {\it Herschel}:
it still has to be used for observations of bright Galactic fields and to be unset for
faint Galactic fields or extragalactic targets, but its role is limited to defining
whether a source mask will be automatically built and used during baseline subtraction
and average drift subtraction to protect bright sources, or not. If the {\tt /galactic}
option is set when it should not, then the source mask may be heavily contaminated
by noise, and in that case the large-scale drift will be poorly subtracted.
Conversely, if the {\tt /galactic} option is forgotten when processing a mosaic of a
Galactic star formation region, then the baselines will be contaminated by sources,
and diffuse emission may be suppressed to a large extent (depending on the relative
geometry of scans and sources). No quantitative brightness threshold can be given,
since such a threshold would largely depend on observing conditions. However, when the
{\tt /galactic} option is used, the final source mask is saved as the last plane of the
FITS cube output (see Section \ref{output} below) and can be inspected. If it is found that
diffuse regions were abusively masked, then it is recommended to restart the processing
with {\tt galactic = 0}\,. \\

The pipeline projects the data at the end of the processing, but final maps are also
produced by {\it Scanamorphos}, using a slightly less accurate projection matrix,
in the sense that the imprint of each bolometer is approximated by a disk instead of
a square. We here discuss only the latter maps. The choice of the pixel size and map orientation
is left to the user. The pixel size can be different from the default if, for example, it is
wished to produce congruent maps at different wavelengths. However, not all choices will result
in adequate coverage of the field of view or good sampling of the point spread function.
It may also be desirable to build maps in scan coordinates rather than right
ascension and declination, for very elongated fields of view for instance.

The user has control over the inclusion of turnaround data (at field of view
edges, acquired while the telescope acceleration is non-zero). Turnaround data
are always used in the drifts determination, whenever they are included in the
input data, but are weighted differently to prevent artificially high coverage values;
they can however be excluded before the final projection.

\clearpage

\section{Memory usage control}

For very large data volumes, two parameters controlling memory usage can be modified
directly inside the main program. If you wish to explore this possibility, it is best
to ask for advice by sending a message to the developer.

The average drift subtraction is the most demanding processing step in terms of memory.
If the memory usage for that step exceeds about 60-70\% of the machine capacity (for deep
observations of small fields or shallow observations of very wide fields), then the processing
may become prohibitively long. In that case, it is advised to momentarily increase the stability
length used specifically for that step. This is equivalent to increasing the minimum timescale
of the average drift, which controls the size of the drift difference matrix. This is
achieved by decreasing the {\tt max\_nt\_coarse} parameter in the main program.
The default value corresponds to a $0.40$~s timescale for the average drift
of the full mosaic of NGC\,6334, instead of $0.18$~s for the instrumental drifts.

The {\tt max\_n\_samples} parameter controls whether
the field of view has to be sliced into several overlapping spatial blocks, that are
processed separately and stitched together at the end of the processing; this is useful
when the data volume is too large to be treated as a single block, but a drawback is that
the separation between the blocks is made automatically and may not result in an
optimal location of the edge between two maps. To prevent slicing the field into
several blocks, the {\tt max\_n\_samples} parameter has to be increased.

A command-line option explicitely allows the number of spatial blocks to be specified.
This option was introduced to allow the time resolution for the average drift computation
to be increased (see Section\,3.8 of the companion paper), and may be useful for some
ArT\'eMiS mosaics with very little redundancy at joints between subregions, provided
these subregions coincide more or less with the automatically-defined blocks.

\clearpage

\section{Non-default astrometry}

It is possible to specify astrometric offsets for each scan (angular distances
in arcsec) with the {\tt offset\_ra\_as} and {\tt offset\_dec\_as} parameters.
These offsets can be determined for instance from maps projected separately
for each scan, provided they contain relatively bright compact sources.
Pointing offsets found in the input pipeline structures, if they exist,  will already
have been applied during the formatting of the data. \\

By default, the astrometry and spatial grid are determined from the input scans.
The user has however the latitude to change the spatial grid or request that only
a portion of the entire field be processed and mapped: \\
1) by supplying a reference fits header (in the form of an IDL string array).
In that case, the astrometry is taken entirely from the header. In batch mode,
the header is expected to be in the form of an IDL save file, containing
a variable named exactly~ {\tt hdr\_ref}\,. \\
2) by supplying a three-element array containing the central coordinates and minimum
radius of a subfield to be excised from the data (all in degrees):
\begin{verbatim}
  cutout=[ra_center, dec_center, radius]
\end{verbatim}

In case 1, if the reference image from which the header has to be extracted has been produced
by {\it Scanamorphos} (as the first plane of the fits cube), and if it has to be rotated first
by a given angle, here is the sequence of IDL commands to create the reference header~
{\tt hdr\_ref}\,:
\begin{verbatim}
  cube = readfits('ref_ima.fits', hdr_cube)
  hdr_ima = hdr_cube
  sxaddpar, hdr_ima, 'NAXIS', 2
  sxdelpar, hdr_ima, 'NAXIS3'
  hrot, cube(*,*,0), hdr_ima, ima_rot, hdr_ref, angle, -1, -1, 1
\end{verbatim}
where {\it angle} is the value of the rotation angle in degrees, clockwise.
To save the reference header to disk, this command can be used:
\begin{verbatim}
  save, filename='hdr_ref.xdr', hdr_ref, /xdr
     
\end{verbatim}

\section{Output}
\label{output}

As the successive processing steps are run, some information is printed to the terminal
window (and to a log file if the~ {\tt batch\_process\_artemis.pro}~ wrapper is used). In particular,
before the final projection, the processing time and drift amplitudes are indicated.
For those who are familiar with IDL, regular stops in the processing occurring in
visualization mode enable the user to interrupt the code and temporarily switch
to the command line, in order to check the content of variables or re-display images
and graphics with different dynamic ranges. Altering variables is not recommended,
unless a deep knowledge of the data and algorithm have been gained beforehand. \\

\begin{figure}[ht!]
\vspace*{0.5cm}
\centering
\begin{minipage}[!ht]{7.cm}
\includegraphics[width=1.\textwidth,clip]{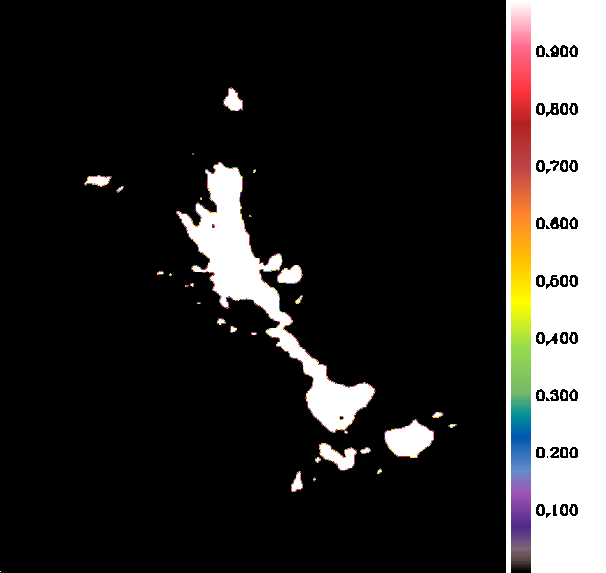}
\end{minipage}
\begin{minipage}[!ht]{7.cm}
\includegraphics[width=1.\textwidth,clip]{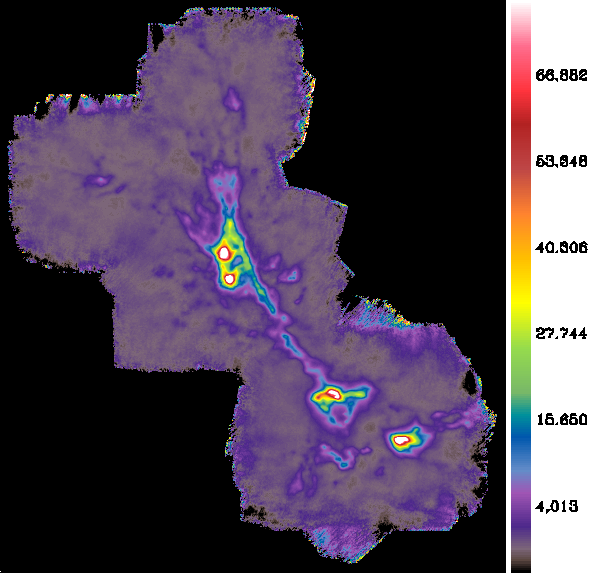}
\end{minipage} \\
\caption{{\bf Left:} Automatically-built source mask. {\bf Right:} Final map projected on the same grid,
in the standard astronomical orientation (RA to the left, N to the top).}
\label{fig:mask}
\end{figure}

On output, the signal, error, drifts and weight maps are assembled into a cube, of
which the third dimension is the plane index. If the~ {\tt /galactic}~ option was used,
the last plane of the cube is the map showing positions included in the source mask built
during the baseline subtraction. This cube and the associated astrometry are saved on disk
in a file conforming to the Flexible Image Transport System (FITS) standard
\citep{Wells81, Hanisch01}.
The weight map should always be inspected, since it allows to select the part of the
signal map with adequate coverage. The source mask should help assessing whether the~
{\tt /galactic}~ option was appropriate or not. Figure~\ref{fig:mask} shows the source mask
produced for NGC\,6334. \\

The processed time series are also reinjected into the pipeline structures (one for each
scan leg), in the~ {\tt donnees\_red}~ field, after appropriate accounting of the bolometers
and time steps discarded during the initial formatting of the data. This allows the
production of the final map with the pipeline. For this to be possible, the data have
to be stored in a directory where the user is allowed to write files. Otherwise, only
the FITS file containing the maps produced by {\it Scanamorphos} can be saved.

\clearpage

\section{To combine overlapping fields}

Overlapping fields can be combined either during the processing or after the final maps
have been created. If the area of overlap is a large fraction of the total area, then the
processing is standard. \\

\noindent
Observations consisting of largely disjoint sub-mosaics may be better processed separately
(which will also be more efficient regarding time and memory usage). The final maps can then
be combined with the~ {\tt stitch\_blocks.pro}~ routine, provided they have been created with
the same reference header and have been renamed appropriately. Care must also be taken that
they are arranged in the optimal order, such that the overlap between successive maps is maximal.
For example, if we wish to use the root name~ {\tt field}~ for~ {\tt n = 3}~ maps, we will rename them:
\begin{verbatim}
  field_0.fits, field_1.fits, field_2.fits
\end{verbatim}
and they are combined with these commands:
\begin{verbatim}
  hdr_ref = headfits('field_0.fits')
  stitch_blocks, nblocks=n, root_out='field', hdr=hdr_ref
\end{verbatim}
which will produce a file named \begin{verbatim}
  field_combined.fits .
     
\end{verbatim}

\section{Notes on memory and CPU requirements}

Development and tests were made on a machine previously dedicated to {\it Herschel} data processing, with 256\,Gb
of memory and 8 cores at 3.5\,GHz. \\

\noindent
To process a field of about $18\arcmin$ on a side at 350\,$\mu$m (with 7 valid subarrays) takes on the order of
1.8 times the on-target observation duration when combining 2 scans, and 2.1 when combining 6 scans (with the full time
resolution of 0.18\,s for the drifts, i.e. a stability length of 0.7 times the FWHM). The processing time thus
increases in a mildly non-linear way with the data volume for a given map geometry. For such a dataset, the total
memory requirement should closely approach 32\,Gb. \\

\noindent
To process the 350\,$\mu$m mosaic of NGC\,6334 (0.17 square degree with very inhomogeneous coverage),
comprising 35 scans, takes almost 2.1 times the observation duration, and the memory requirement is large.
The average drift subtraction on small timescales uses up more than half the available 256\,Gb of memory. \\

\noindent
In case the machine capacities are exceeded, i.e. if you see the error message {\it ``cannot allocate memory''} or
if memory swapping slows down the processing too much, it is recommended to slice the field of view or the observations
into several blocks, either automatically with the~ {\tt nblocks}~ parameter, or manually by selecting subsets of
all the scans, to be processed separately.

\clearpage

\section{Standard usage}

Because {\it Scanamorphos} has been developed as an interactive tool, by default it queries
information about the desired output (``inputs at the prompt'' in Table \ref{tab_options}).
If this is not wanted, then the~ {\tt /batch}~ option should be used, and non-default parameters
can then be supplied on the command line (see the last part of Table \ref{tab_options}). The parameters
discussed here pertain only to {\it Scanamorphos}, and not to the maps produced by the pipeline. \\

Below is a summary of the IDL commands needed to produce the 350\,$\mu$m mosaic of NGC\,6334\,,
after the raw data have been calibrated and corrected for opacities (see the pipeline manual
for these steps). To cleanly interface with the pipeline, the only external routine called
directly by {\it Scanamorphos} is~ {\tt get\_calib\_project\_info.pro}\,, calling itself all the necessary
configuration procedures. {\bf The project name and the applicable calibration table must have
been set correctly}, as explained in the pipeline manual.
{\bf Data for separate arrays or separate targets must be stored in different directories}
(~{\tt dir\_out}~ below). \\

\begin{verbatim}
;;; FORMATTING OF THE INPUT DATA:
array = 350
dir_out = [..directory where input structures, temporary files and output maps will be written..]

;;; with project name  "T-091.F-0008-2013"  and 2013 calibration table:
list_scannum = [43300 + [49, 51, 53, 56, 58, 59, 62, 63, 65, 67], $
                43600 + [42, 47, 52, 54, 58], 57931 + indgen(3), 58030 + [5, 8, 9]]
format_input_scanam_artemis, dir_out=dir_out, list_scannum=list_scannum, array=array

;;; with project name  "E-091.C-0870A-2013"  and 2014 calibration table:
list_scannum = [30200 + [71, 81, 83, 84, 86, 88], 30305, $
   31200 + [52, 56, 73, 79], 33630 + [4, 5, 7]]
format_input_scanam_artemis, dir_out=dir_out, list_scannum=list_scannum, array=array, /append
;;; "/append"  KEYWORD MANDATORY to preserve the earlier contents of  "scanlist_artemis"

;;; Here it's possible to close the session and do the processing later.
;;; The "scanlist_artemis" file written in dir_out can be edited
;;; to remove scans and process only a fraction of the mosaic
;;; (in that case, be careful with the scan selection).

;;; Once the formatting is done, the correct project name is retrieved from the input structures
;;; and no longer needs to be read from the pipeline configuration procedures.


;;; PROCESSING:
file_out = 'n6334_35scans'
pixsize = 2.
nzadata = 'no'
scanam_artemis, /galactic, /batch, array=array, dir_scanlist=dir_out, $
   file_out=file_out, pixsize=pixsize, nzadata=nzadata
      
\end{verbatim}

\noindent
or alternatively, to have an interactive run and visualize the scan trajectories: \\
\begin{verbatim}
scanam_artemis, /galactic, /visu, /vis_traject, dir_scanlist=dir_out
   
\end{verbatim}

\noindent
Note that the input~ {\tt "array=350"}~ is not needed, because by default the 350\,$\mu$m band is assumed.
The wavelength does not need to be specified in the call to~ {\tt scanam\_artemis}\,, since it
is retrieved from the input structures, but if equal to 450\,$\mu$m it is mandatory to include~
{\tt "array=450"}~ in the call to~ {\tt format\_input\_scanam\_artemis}\,.
The pixel size of the final map is not needed unless it differs from the default value (FWHM $/ 4$). \\

To produce a log file containing the messages displayed during the processing (saved in the
same directory as the input structures and output maps), one may also use
the~ {\tt batch\_process\_artemis.pro}~ wrapper: \\
\begin{verbatim}
batch_process_artemis, master_list='[..path_to../]master_list_n6334'
   
\end{verbatim}
where the~ {\tt master\_list}~ ascii file contains the information needed to retrieve the data
and processing parameters of one or several observations. A couple of examples are given in
the~ {\tt example/}~ subdirectory of the distribution. Please note that this way
of calling~ {\tt scanam\_artemis}~ does not allow to obtain full error messages and thus to track
the origin of potential coding errors.

\begin{table*}[!h]
\vspace*{1cm}
\caption{Summary of inputs and options to~ {\tt scanam\_artemis}}
\label{tab_options}
\centering
\begin{tabular}{ll}
\hline\hline
\multicolumn{2}{c}{auxiliary file} \\
\hline
{\tt scanlist\_artemis} & ascii file containing the directory and file names of the input scans \\
~                       & (created automatically by~ {\tt format\_input\_scanam\_artemis}\,) \\
\hline\hline
\multicolumn{2}{c}{command line keywords and parameters} \\
\hline
{\tt dir\_scanlist}     & directory where the~ {\tt scanlist\_artemis}~ file and input structures are stored \\
{\tt visu}              & select to visualize intermediate results and stop after each major step \\
{\tt vis\_traject}      & select to visualize the imprint and orientation of all scans \\
{\tt debug}             & select for detailed visualizations (for developers only) \\
{\tt galactic}          & select to automatically build a mask source during baseline subtraction \\
{\tt noglitch}          & select to bypass glitch masking \\
{\tt hdr\_ref}          & reference header to enforce the same astrometry~\tablefootmark{a} \\
~                       & (IDL string array or IDL save file containing a variable named~ {\tt hdr\_ref}\,) \\
{\tt cutout}            & central RA and DEC coordinates and radius of the subfield to be processed and mapped \\
{\tt nblocks}           & number of spatial blocks the field of view will be sliced into~\tablefootmark{b} \\
{\tt block\_start}      & index of the first spatial block to be processed, when nblocks $> 1$~\tablefootmark{b} \\
{\tt one\_plane\_fits}  & select to save each plane into a separate fits file \\
\hline\hline
\multicolumn{2}{c}{inputs at the prompt} \\
\hline
output directory        & directory where final maps and intermediate variables will be stored~\tablefootmark{c} \\
~                       & (default: directory containing the input structures) \\
output FITS file root   & full file name = file root + instrument + wavelength + block index + ``.fits'' \\
~                       & (default: target name as it appears in the metadata) \\
map orientation         & binary choice: orientation of the first scan, or standard astronomical orientation (default) \\
pixel size              & in arcsec, equal to FWHM $/ 4$ by default \\
non-zero acceleration data   & option to reject turnaround data from final map (they are included by default) \\
\hline\hline
\multicolumn{2}{c}{parameters in batch mode} \\
\hline
{\tt batch}             & select to avoid typing inputs at the prompt \\
{\tt file\_out}         & same as ``output FITS file root'' above \\
{\tt array}             & either~ {\tt 350}~ (350\,$\mu$m, default) or~ {\tt 450}~ (450\,$\mu$m) \\
{\tt orient}            & same as ``map orientation'' above: either~ {\tt "scan"}~ or~ {\tt "astro"}~ (default) \\
{\tt pixsize}           & same as ``pixel size'' above \\
{\tt nzadata}           & {\tt "no"}~ to reject turnaround data or~ {\tt "yes"}~ to include them (default) \\
\hline
\end{tabular}
\tablefoot{
\tablefoottext{a}{The field of view of the reference header must of course
be the same as that of the input scans.}
\tablefoottext{b}{The option to slice the field of view should be used only in very specific cases:
see Section\,3.8 of the companion paper.
When the first blocks have already been processed, if the program is interrupted, it is
possible to restart from the first unprocessed block with the~ {\tt block\_start}~ parameter.
}
\tablefoottext{c}{Make sure to have enough space in this directory to store all the intermediate
variables necessary for the processing, and to be able to write files in it.
}
}
\vspace*{1cm}
\end{table*}
~ \\

\vfill
\noindent
{\bf The software is distributed in the hope that it will be useful in a broad sense to ESO and OSO submillimeter
observers and researchers planning to reprocess public ArT\'eMiS data. Users of~ {\tt scanam\_artemis}~ are thus
encouraged to contact the developer regarding bug reports, the documentation, or any relevant query of general interest.
Questions regarding the pipeline or the instrument should be directed to the instrument team.}

\end{document}